\documentclass[10pt]{article}
\usepackage{multicol}
\usepackage{graphicx}
\usepackage{amsmath}
\usepackage[a4paper]{geometry}
\usepackage{hyperref}
\usepackage{rotating}

\begin{document}

\begin{center}
{\Large \bf An analysis of transverse momentum spectra of various
jets produced in high energy collisions}

\vskip.75cm

Yang-Ming Tai$^{1,2,}${\footnote{E-mail: taiyangming6@qq.com}},
Pei-Pin Yang$^{1,2,}${\footnote{E-mail: yangpeipin@qq.com;
peipinyangshanxi@163.com}}, Fu-Hu
Liu$^{1,2,}${\footnote{Corresponding author. E-mail:
fuhuliu@163.com; fuhuliu@sxu.edu.cn }}
\\

{\small\it $^1$Institute of Theoretical Physics \& State Key
Laboratory of Quantum Optics and Quantum Optics Devices,\\ Shanxi
University, Taiyuan, Shanxi 030006, People's Republic of China

$^2$Collaborative Innovation Center of Extreme Optics, Shanxi
University, \\ Taiyuan, Shanxi 030006, People's Republic of China}

\end{center}

\vskip.5cm

{\bf Abstract:} With the framework of the multi-source thermal
model, we analyze the experimental transverse momentum spectra of
various jets produced in different collisions at high energies.
Two energy sources, a projectile participant quark and a target
participant quark, are considered. Each energy source (each
participant quark) is assumed to contribute to the transverse
momentum distribution to be the TP-like function, i.e. a revised
Tsallis--Pareto-type function. The contribution of the two
participant quarks to the transverse momentum distribution is then
the convolution of two TP-like functions. The model distribution
can be used to fit the experimental spectra measured by different
collaborations. The related parameters such as the entropy
index-related, effective temperature, and revised index are then
obtained. The trends of these parameters are useful to understand
the characteristic of high energy collisions.
\\

{\bf Keywords:} Transverse momentum spectra, High-energy jets,
TP-like function
\\

{\bf PACS:} 12.40.Ee, 13.87.Fh, 25.75.Ag, 24.10.Pa

\vskip1.0cm

\begin{multicols}{2}

{\section{Introduction}}

In central heavy ion (nucleus-nucleus, A-A) collisions at high
energy, quark-gluon plasma (QGP) is believed to create
possibly~\cite{1,2,3}, because the environment of high temperature
and density is formed. After the formation, QGP experiences the
process of hadronization and then produces lots of final-state
particles. Meanwhile, at the early stage of collisions, some
products such as various jets are produced and interact
subsequently with QGP. Because of the interactions between jets
and QGP, jets lost their energies when they go through QGP region.
Not only lots of identified particles but also various jets can be
measured in experiments at high energies. Indeed, in the abundant
data on high energy collisions, the data on various jets are one
of the most important constituents. We are interested in analyzing
the experimental transverse momentum ($p_T$) spectra of various
jets, because they can reflect some information of early
collisions of participant quarks or partons.

Generally, the $p_T$ spectra of various jets are wider than those
of identified particles. In fact, both the $p_T$ spectra of
various jets and identified particles cover a wide $p_T$ range.
Even if for the later, one may divide the $p_T$ range into low-
and high-$p_T$ regions. It is expected that the spectra in
low-$p_T$ region are contributed by the soft excitation process,
while the spectra in high-$p_T$ region are contributed by the hard
scattering process. In some cases, the spectra in low- and
high-$p_T$ regions are still complex. One may divide further the
low- and high-$p_T$ regions into very low- and low-$p_T$ regions
as well as high- and very high-$p_T$ regions respectively. It is
expected that the spectra in different $p_T$ regions can be
analyzed by different functions. This means that one needs
two-component or even four-component function to fit the wide
$p_T$ spectra.

It is known that perturbative Quantum Chromodynamics (pQCD)
successfully describes the processes which involve large momentum
transfers. In particular for proton-proton (p-p) collisions,
multi-jet production at high-$p_T$ is well described if initial-
and final-state radiations are considered (see e.g.
ref.~\cite{3a}). In addition, multi-parton interactions may play a
role at low-$p_T$~\cite{3b}. However, pQCD is very complex, which
limits its wider applications in high energy proton-nucleus (p-A)
and A-A collisions. We hope to use an alternative and thermal-like
or statistical method to describe uniformly the spectra of various
jets and identified particles in both low- and high-$p_T$ regions
in p-p, p-A, and A-A collisions at thigh energies. As the first
step and as an example, we consider the two-component function.

There are two methods to superpose the two components in a
function~\cite{4,5,6}. The first method uses a weighted sum for
the two components and there are correlations between the
parameters of the two components, though the point of linkage is
smooth. The second method uses a step function to link the two
components~\cite{6} and there is a non-smooth linkage between the
two components, though the parameters are uncorrelative. It is
imaginable that more issues will appear if we consider four
components in a function. Although the two-component function is
widely used in literature, it is not an ideal treatment method,
not to mention the four-component function. We hope to use a
method to treat the two or four components uniformly. Even a
single component function is used to fit the spectra in wide $p_T$
range.

Fortunately, to search for the single component function for the
spectra in wide $p_T$ range is possible, because the similarity,
universality, or common law is existent in high energy
collisions~\cite{7,8,9,10,11,11a,11b}. To search for the single
component function, we have tested many potential functions.
Finally, we have found that the convolution of two or more revised
Tsallis--Pareto-type functions~\cite{12,13} is a suitable choice.
For the purpose of doing a convenient description, we call the
revised Tsallis--Pareto-type function the TP-like function in our
recent work~\cite{13} and this paper. In fact, the
Tsallis--Pareto-type function is more proper to restrict only to a
Tsallis distribution, once one essentially uses the non-extensive
statistical mechanics~\cite{13a}. Because the Tsallis distribution
has more than one forms of expression, the term of
Tsallis--Pareto-type function is used in our work to mention the
concrete form.

The application of the convolution of two or more functions is a
general treatment method with the framework of the multi-source
thermal model~\cite{5}, where the considered distributions are
assumed from the contributions of two or more energy sources. The
considered distributions include at least the multiplicity,
transverse energy, and transverse momentum (transverse mass)
distributions. At least two energy sources are considered in the
collisions. Three or more energy sources are not excluded, if the
two energy sources are not enough to fit the spectra. The concrete
number of energy sources is determined by the quality of fits and
the scenario of physics. This fit methodology is suitable and
unified for the analyses of spectra in different rapidity
intervals, centrality classes, and collision systems.

In this paper, in the framework of the multi-source thermal model,
we assume that a projectile participant quark and a target
participant quark take part in the production of various jets, and
they contribute to the $p_T$ distribution to be the TP-like
function~\cite{12,13}. Then, we may use the convolution of two
TP-like functions to fit the experimental $p_T$ spectra of various
jets. The related data quoted in this paper are from
proton-(anti)proton (p-p($\rm\bar p$)), deuteron-gold (d-Au),
gold-gold (Au-Au), proton-lead (p-Pb), and lead-lead (Pb-Pb)
collisions, with different selection conditions, over a
center-of-mass energy ($\sqrt{s_{NN}}$, or simplified as
$\sqrt{s}$ for p-p($\rm\bar p$) collisions) range from 0.2 to 13
TeV.

The remainder of this paper is structured in the following. The
formalism and method are described in Section 2. The results and
discussion are given in Section 3. Finally, we give the summary
and conclusions in Section 4.
\\

{\section{The formalism and method}}

According to ref.~\cite{12}, the Tsallis--Pareto-type function
which describes empirically the $p_T$ spectra of particles with
rest mass $m_0$ can be given by
\begin{align}
f_{p_T}(p_T)= Cp_T\left(1+ \frac{\sqrt{p_T^2+m_0^2}-
m_0}{nT}\right)^{-n}
\end{align}
which is a probability density function and $C$ is the
normalization constant because $\int_0^{\infty}
f_{p_T}(p_T)dp_T=1$. In Eq. (1), as an entropy index-related
parameter, $n$ is related to the entropy index $q$ because
$n=1/(q-1)$. Generally, $q=1$ or $n=\infty$ means an equilibrium
state. If $q$ is close to 1 or $n$ is large enough, the system is
close to an equilibrium state. The free parameter $T$ in Eq. (1)
is an effective temperature that describes the excitation and
expansion degree of the emission source for particles. We call $T$
the effective temperature because both the contributions of random
thermal motion and flow effect are included.

Equation (1) is not flexible enough in the description of $p_T$
spectra of particles, in particular for the spectra in low-$p_T$
region. Empirically, Eq. (1) can be revised artificially by adding
a revised index $a_0$ that is non-dimensional as the power index
of $p_T$. Then, we have the TP-like function to be~\cite{13}
\begin{align}
f_{p_T}(p_T)= Cp_T^{a_0}\left( 1+
\frac{\sqrt{p_T^2+m_0^2}-m_0}{nT} \right)^{-n},
\end{align}
where $C$ is the normalization constant which is different from
that in Eq. (1). For the purpose of convenience, two normalization
constants in Eqs. (1) and (2) are represented by the same symbol
$C$, though they may be different. Although one more parameter is
introduced, Eq. (2) is more accurate than Eq. (1). In particular,
we can obtain Eq. (1) from Eq. (2) if we use $a_0=1$.

With the framework of the multi-source thermal model~\cite{5}, we
assume that many quarks or partons take part in the collisions.
Each quark or parton is regarded as an energy source. For a given
particle or jet, two quarks, i.e. a projectile participant (the
first) quark and a target participant (the second) quark, play
main role in the production process. Other quarks that take part
in the interactions with weak contributions can be neglected. For
the two main quarks, the contribution amount or portion ($p_{ti}$)
of each quark to $p_T$ is assumed to obey the TP-like function,
where $i=1$ and 2 are for the first and second quarks
respectively. The TP-like function obeyed by $p_{ti}$ is~\cite{13}
\begin{align}
f_i(p_{ti})= C_ip_{ti}^{a_0} \left( 1+
\frac{\sqrt{p_{ti}^2+m_{0i}^2}-m_{0i}}{nT} \right)^{-n},
\end{align}
where $m_{0i}$ is empirically the constituent mass of the $i$-th
participant quark.

The total amount contributed by the two quarks is the convolution
of two TP-like functions. That is~\cite{13}
\begin{align}
f_{p_T}(p_T)=\int_0^{p_T} f_1(p_{t1})f_2(p_T-p_{t1})dp_{t1}
\end{align}
or
\begin{align}
f_{p_T}(p_T)=\int_0^{p_T} f_2(p_{t2})f_1(p_T-p_{t2})dp_{t2},
\end{align}
where the functions $f_1$ and $f_2$ are given by Eq. (3) for
various jets which are produced by different collisions which are
listed in the table in the next section. In most cases, the
convolution of two TP-like functions is suitable for the spectra
of various jets. Correspondingly, two heavy flavor quarks such as
$c+\bar c$, $b+\bar b$, or $t+\bar t$ should be considered due to
more effective energy being needed. For two light flavor quarks
such as $u+\bar u$, $d+\bar d$, or $s+\bar s$, we do not need to
consider them due to too less effective energy for the production
of various jets.

The method of the convolution of three TP-like functions is
similar to that of two TP-like functions. Firstly, we may obtain
the convolution $f_{12}(p_{t12})$ of the first two TP-like
functions $f_{1}(p_{t1})$ and $f_2(p_{t2})$. Secondly, we may
obtain the convolution $f_{p_T}(p_T)$ of $f_{12}(p_{t12})$ and
$f_3(p_{t3})$. Alternatively, we may obtain firstly the
convolution $f_{23}(p_{t23})$ of the last two TP-like functions
$f_{2}(p_{t2})$ and $f_3(p_{t3})$, and then we may obtain the
convolution $f_{p_T}(p_T)$ of $f_1(p_{t1})$ and $f_{23}(p_{t23})$.
The same idea can be used for the convolution of more than three
TP-like functions. At present, the convolution of a projectile
participant quark and a target participant quark is enough to fit
the spectra of $p_T$ of various jets. Temporarily, we do not need
to consider the convolution of three or more participant quarks.

Because of the introduction of $a_0$, Eq. (2) is more accurate and
flexible than Eq. (1). By using $a_0$, the spectra in very
low-$p_T$ region can be described reasonable. With the framework
of the multi-source thermal model, the method of the convolution
of two or more probability density functions is applicable for not
only the spectra of $p_T$ but also the spectra of multiplicity and
transverse energy. In our analysis, the free parameters are $n$,
$T$, and $a_0$. The normalization constant is a parameter, but not
a free parameter. The convolution does not introduce new free
parameters, but the source number from the collision picture. In
the method, to search for the probability density function
contributed by a single participant or contributor or source is a
key issue. This participant or contributor or source can be quark
if we study the spectra of particles, or nucleon if we study the
spectra of nuclear fragments.
\\

\begin{figure*}[!htb]
\begin{center}
\includegraphics[width=15.0cm]{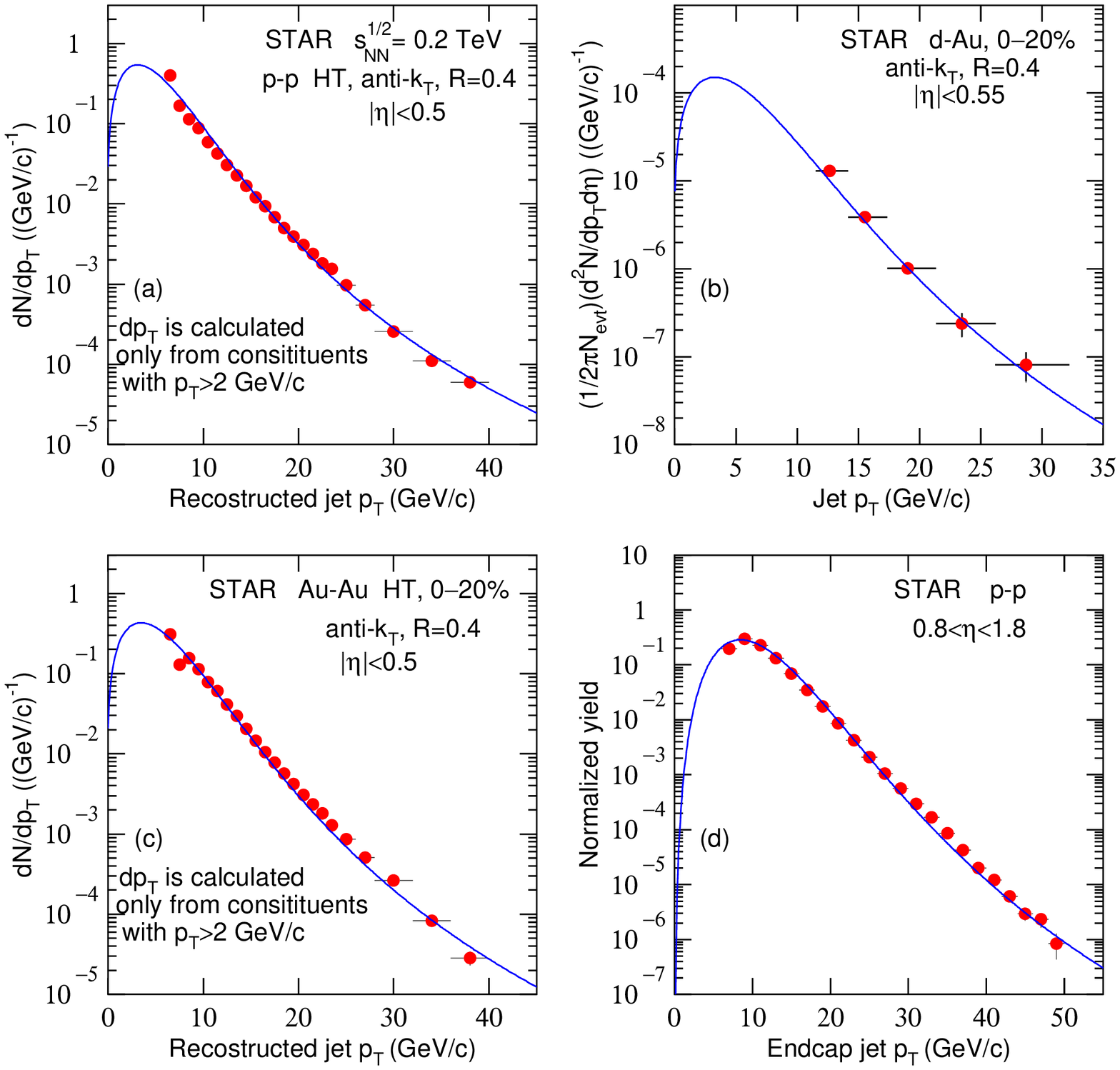}
\end{center}
{\small Figure 1. Transverse momentum spectra of different jets
produced in (a) p-p, (b) d-Au, and (c) Au-Au collisions with
mid-$\eta$ ($|\eta|<0.5$ or 0.55), as well as in (d) p-p
collisions with non-mid-$\eta$ ($0.8<\eta<1.8$) at
$\sqrt{s_{NN}}=0.2$ TeV. The symbols are cited from the
experimental data measured by the STAR
Collaboration~\cite{14,15,16} and the curves are our fitted
results with Eq. (4).}
\end{figure*}

\begin{table*}[!htb]
{\small Table 1. Values of $n$, $T$, $a_0$, $N_0$ ($\sigma_0$),
$\chi^2$, and ndof corresponding to the curves in Figures 1--9,
where $\sigma_0$ is only for Figure 2(a). \vspace{-.75cm}}
{\footnotesize
\begin{center}
\newcommand{\tabincell}[2]{\begin{tabular}{@{}#1@{}}#2\end{tabular}}
\begin{tabular} {ccccccccccccc}\\ \hline\hline
Figure and type & Collision & $n$ & $T$ (GeV) & $a_{0}$ & $N_0$ [$\sigma_0$ (mb)] & $\chi^2$/ndof \\
\hline
Figure 1(a), p-p, $|\eta|<0.5$   & $t+\bar t$ & $2.80\pm0.12 $ & $0.032\pm0.008$  & $0.00\pm0.01$  & $3.499\pm0.002$  & 225/19\\
Figure 1(b), d-Au, $|\eta|<0.55$ & $t+\bar t$ & $3.20\pm0.02 $ & $0.036\pm0.008$  & $0.00\pm0.01$  & $0.007\pm0.001$  & 5/1\\
Figure 1(c), Au-Au, $|\eta|<0.5$ & $t+\bar t$ & $3.20\pm0.05 $ & $0.040\pm0.009$  & $0.00\pm0.05$  & $2.999\pm0.002$  & 249/19\\
Figure 1(d), p-p, $0.8<\eta<1.8$ & $t+\bar t$ & $6.00\pm0.05 $ & $0.045\pm0.001$  & $1.50\pm0.01$  & $2.499\pm0.002$  & 195/18\\
\hline
Figure 2(a), p-Pb, 0$-$20\%      & $t+\bar t$ & $3.00\pm0.01 $ & $0.130\pm0.002$  & $1.00\pm0.02$  & $0.999\pm0.001$  & 18/4\\
Figure 2(a), p-Pb, 20$-$40\%     & $t+\bar t$ & $3.00\pm0.02 $ & $0.130\pm0.001$  & $1.00\pm0.02$  & $0.699\pm0.002$  & 11/3\\
Figure 2(a), p-Pb, 40$-$60\%     & $t+\bar t$ & $3.00\pm0.01 $ & $0.130\pm0.003$  & $1.00\pm0.02$  & $0.499\pm0.001$  & 12/2\\
Figure 2(a), p-Pb, 60$-$80\%     & $t+\bar t$ & $3.00\pm0.03 $ & $0.130\pm0.001$  & $1.00\pm0.02$  & $0.299\pm0.002$  & 18/2\\
Figure 2(a), p-Pb, 80$-$100\%    & $t+\bar t$ & $3.00\pm0.05 $ & $0.130\pm0.001$  & $1.00\pm0.02$  & $0.149\pm0.002$  & 6/1\\
Figure 2(b), Pb-Pb, 0$-$10\%     & $t+\bar t$ & $3.00\pm0.11 $ & $0.130\pm0.050$  & $1.00\pm0.01$  & $(1.799\pm0.001)\times10^{-5}$ & 8/5\\
Figure 2(b), Pb-Pb, 10$-$30\%    & $t+\bar t$ & $3.00\pm0.05 $ & $0.140\pm0.051$  & $1.00\pm0.01$  & $(2.299\pm0.001)\times10^{-5}$ & 3/5\\
Figure 2(b), Pb-Pb, 30$-$50\%    & $t+\bar t$ & $3.00\pm0.21 $ & $0.150\pm0.052$  & $1.00\pm0.01$  & $(2.999\pm0.001)\times10^{-5}$ & 4/4\\
Figure 2(b), Pb-Pb, 50$-$80\%    & $t+\bar t$ & $3.00\pm0.31 $ & $0.150\pm0.051$  & $1.00\pm0.01$  & $(3.999\pm0.002)\times10^{-5}$ & 7/3\\
\hline
Figure 3(a), p-${\rm\bar p}$, $Z\rightarrow\mu\mu$ & $t+\bar t$ & $2.10\pm0.02 $ & $1.150\pm0.091$  & $-0.45\pm0.01$ & $259969.508\pm2.200$   & 106/31\\
Figure 3(b), p-${\rm\bar p}$, $Z\rightarrow ee$    & $t+\bar t$ & $2.05\pm0.05 $ & $1.200\pm0.092$  & $-0.45\pm0.01$ & $239964.205\pm2.522$   & 104/31\\
Figure 3(c), p-p, lepton+jet       & $t+\bar t$ & $5.00\pm0.55 $ & $13.000\pm0.853$ & $-0.48\pm0.01$ & $199762.511\pm11.050$  & 257/18\\
Figure 3(c), p-p, dilepton+jet     & $t+\bar t$ & $5.00\pm0.80 $ & $12.000\pm0.852$ & $-0.48\pm0.01$ & $51954.412\pm2.502$    & 201/19\\
Figure 3(d), p-p, lepton+$b$-jet   & $b+\bar b$ & $10.00\pm1.50$ & $12.000\pm0.893$ & $1.00\pm0.01$  & $19997.136\pm0.520$    & 133/29\\
Figure 3(d), p-p, dilepton+$b$-jet & $b+\bar b$ & $10.00\pm1.50$ & $12.000\pm0.955$ & $1.00\pm0.01$  & $2599.627\pm0.223$     & 57/25\\
\hline
Figure 4(a), 7 TeV p-p, $e$+$b$-jet   & $b+\bar b$ & $16.00\pm0.03$ & $15.000\pm1.512$ & $1.00\pm0.03$  & $13999.407\pm0.620$    & 345/12\\
Figure 4(b), 7 TeV p-p, $\mu$+$b$-jet & $b+\bar b$ & $16.00\pm0.05$ & $16.000\pm1.502$ & $1.00\pm0.01$  & $14998.827\pm0.953$    & 230/12\\
Figure 4(c), 7 TeV p-p, $e$+jet       & $t+\bar t$ & $3.40\pm0.15 $ & $3.500\pm0.550$  & $1.60\pm0.13$  & $27400.066\pm0.965$    & 859/20\\
Figure 4(d), 7 TeV p-p, $\mu$+jet     & $t+\bar t$ & $3.50\pm0.12 $ & $3.500\pm0.521$  & $1.50\pm0.10$  & $34924.095\pm0.856$    & 155/20\\
Figure 4(e), 8 TeV p-p, $e$+jet       & $t+\bar t$ & $3.50\pm0.42 $ & $3.000\pm0.552$  & $1.50\pm0.01$  & $119830.499\pm2.505$   & 509/16\\
Figure 4(f), 8 TeV p-p, $\mu$+jet     & $t+\bar t$ & $3.20\pm0.35 $ & $3.000\pm0.512$  & $1.40\pm0.05$  & $159633.552\pm3.254$   & 832/15\\
\hline
Figure 5, leading jet           & $t+\bar t$ & $3.00\pm0.02 $ & $3.600\pm0.195$  & $1.10\pm0.12$  & $43983.362\pm2.562$    & 19/4\\
Figure 5, $2^{nd}$ jet          & $t+\bar t$ & $3.00\pm0.02 $ & $1.000\pm0.190$  & $1.70\pm0.15$  & $41991.806\pm3.255$    & 20/3\\
Figure 5, $3^{rd}$ jet          & $t+\bar t$ & $2.50\pm0.01 $ & $0.600\pm0.185$  & $1.00\pm0.14$  & $42998.320\pm3.235$    & 6/2\\
Figure 5, $4^{th}$ jet          & $t+\bar t$ & $2.50\pm0.01 $ & $0.300\pm0.095$  & $1.00\pm0.16$  & $29999.717\pm2.253$    & 8/1\\
Figure 5, $5^{th}$ jet          & $t+\bar t$ & $2.50\pm0.01 $ & $0.150\pm0.090$  & $1.00\pm0.12$  & $24999.942\pm1.025$    & 7/0\\
\hline
Figure 6(a), leading $b$-jet    & $b+\bar b$ & $9.00\pm0.06 $ & $13.000\pm1.506$ & $1.00\pm0.05$  & $1097.905\pm0.252$     & 98/20\\
Figure 6(b), subleading $b$-jet & $b+\bar b$ & $6.00\pm0.10 $ & $4.800\pm0.502$  & $1.00\pm0.03$  & $1498.243\pm0.586$     & 11/10\\
Figure 6(c), leading jet        & $t+\bar t$ & $2.01\pm0.15 $ & $7.000\pm0.520$  & $-0.48\pm0.01$ & $4998.740\pm1.562$     & 94/14\\
Figure 6(d), subleading jet     & $t+\bar t$ & $2.10\pm0.22 $ & $2.050\pm0.510$  & $-0.48\pm0.01$ & $6965.727\pm0.852$     & 37/14\\
\hline
Figure 7(a), leading light jet    & $c+\bar c$ & $5.00\pm0.18 $ & $11.000\pm1.505$ & $1.00\pm0.02$  & $5629.301\pm0.560$     & 147/36\\
Figure 7(b), subleading light jet & $c+\bar c$ & $5.50\pm0.12 $ & $4.500\pm0.150$  & $1.00\pm0.01$  & $11848.861\pm0.805$    & 89/36\\
Figure 7(c), leading jet          & $t+\bar t$ & $2.50\pm0.15 $ & $5.000\pm0.502$  & $0.00\pm0.03$  & $399824.415\pm5.600$   & 16/32\\
Figure 7(d), subleading jet       & $t+\bar t$ & $2.50\pm0.17 $ & $2.000\pm0.105$  & $0.00\pm0.01$  & $499979.349\pm2.354$   & 28/32\\
\hline
Figure 8(a), small-R $e$+jet      & $t+\bar t$ & $70.00\pm10.10$& $70.000\pm15.100$& $-0.50\pm0.02$ & $39936.022\pm0.500$    & 23/16\\
Figure 8(b), small-R $\mu$+jet    & $t+\bar t$ & $70.00\pm10.05$& $90.000\pm15.520$& $-0.50\pm0.05$ & $34825.665\pm0.850$    & 103/16\\
Figure 8(c), large-R $e$+jet      & $t+\bar t$ & $7.00\pm0.21$  & $25.000\pm1.510$ &  $1.00\pm0.03$ & $174997.093\pm1.220$   & 3/15\\
Figure 8(d), large-R $\mu$+jet    & $t+\bar t$ & $7.00\pm0.20$  & $24.000\pm1.521$ &  $1.00\pm0.03$ & $174997.733\pm1.054$   & 3/15\\
\hline
Figure 9(a), leading, Zjj        & $t+\bar t$ & $3.00\pm0.30$  & $3.200\pm0.201$  &  $1.00\pm0.05$ & $89955.257\pm1.523$    & 230/11\\
Figure 9(b), subleading, Zjj     & $t+\bar t$ & $3.00\pm0.30$  & $1.500\pm0.100$  &  $1.00\pm0.05$ & $99992.944\pm1.554$    & 40/11\\
Figure 9(c), leading, pre-fit    & $t+\bar t$ & $3.00\pm0.11$  & $4.800\pm0.050$  &  $1.00\pm0.02$ & $93136953.879\pm15.400$& 191/9\\
Figure 9(d), fourth, pre-fit     & $t+\bar t$ & $3.50\pm0.10$  & $0.300\pm0.010$  &  $1.00\pm0.02$ & $11991257.185\pm15.600$& 14/6\\
\hline
\end{tabular}
\end{center}}
\end{table*}

\begin{figure*}[!htb]
\begin{center}
\includegraphics[width=15.0cm]{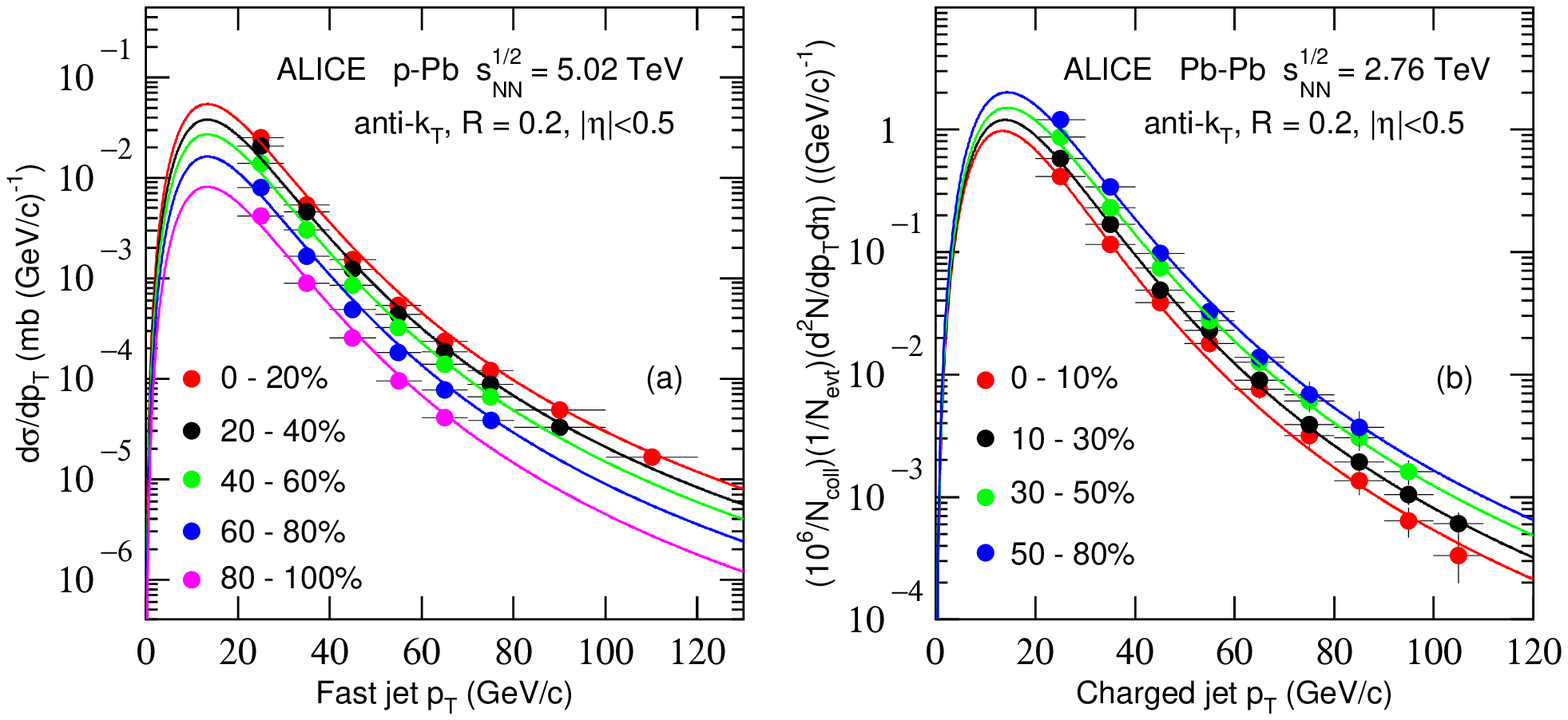}
\end{center}
{\small Figure 2. Transverse momentum spectra of (a) fast jets in
p-Pb collisions at $\sqrt{s_{NN}}=5.02$ TeV and (b) charged jets
in Pb-Pb collisions at $\sqrt{s_{NN}}=2.76$ TeV. The different
symbols are cited from the experimental data with different
centrality classes measured by the ALICE
Collaboration~\cite{18,19} and the curves are our fitted results
with Eq. (4).}
\end{figure*}

\begin{figure*}[!htb]
\begin{center}
\includegraphics[width=15.0cm]{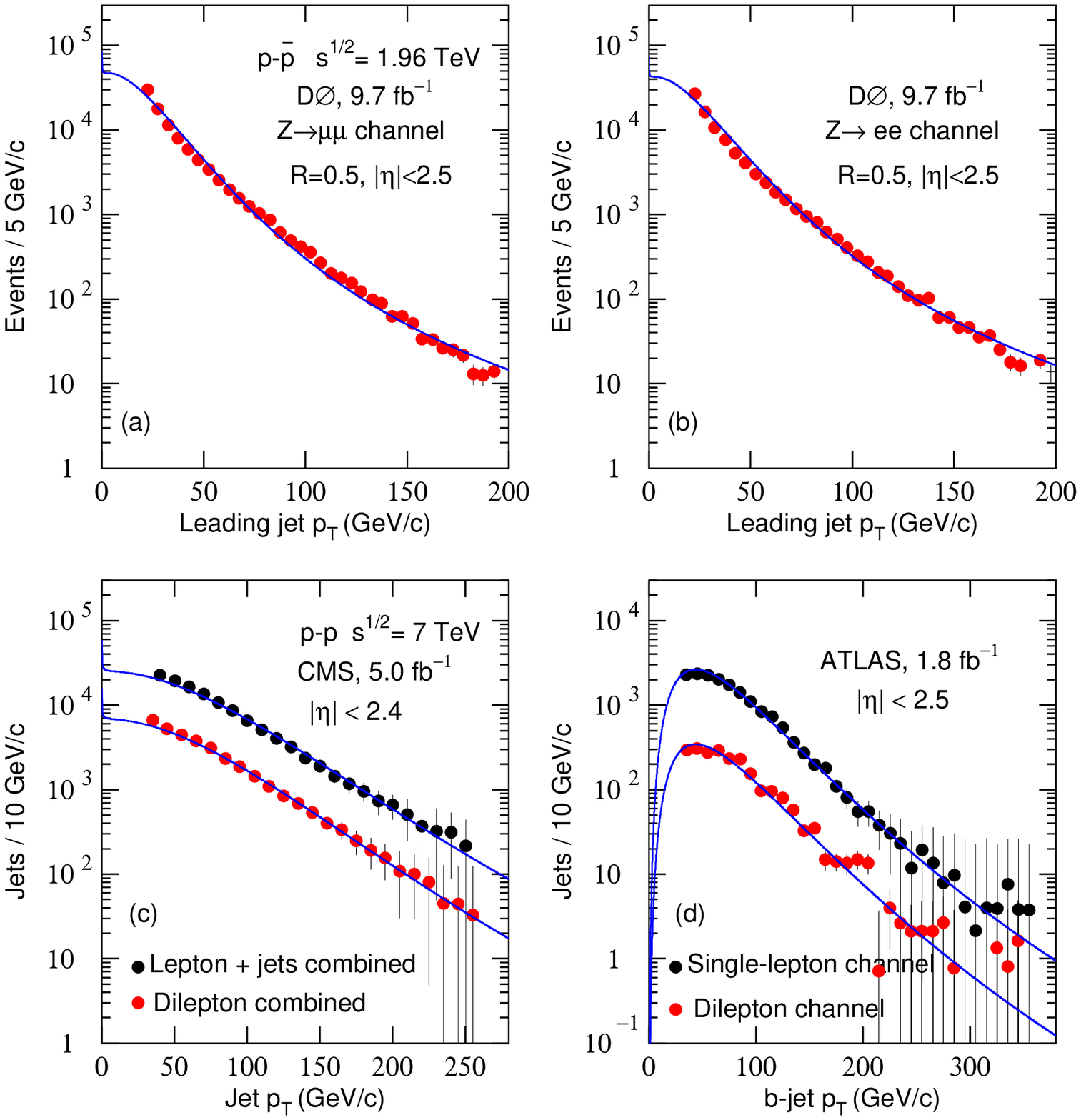}
\end{center}
{\small Figure 3. Transverse momentum spectra of (a)(b) leading
jets with (a) $Z\rightarrow \mu\mu$ and (b) $Z\rightarrow ee$ in
p-$\rm\bar p$ collisions at $\sqrt{s}=1.96$ TeV, as well as (c)
jets and (d) $b$-jets in p-p collisions at $\sqrt{s_{NN}}=7$ TeV
with different channels. The symbols are cited from the
experimental data measured by the D0~\cite{20,21}, CMS~\cite{22},
and ATLAS Collaborations~\cite{23} and the curves are our fitted
results with Eq. (4).}
\end{figure*}

\begin{figure*}[!htb]
\begin{center}
\includegraphics[width=15.0cm]{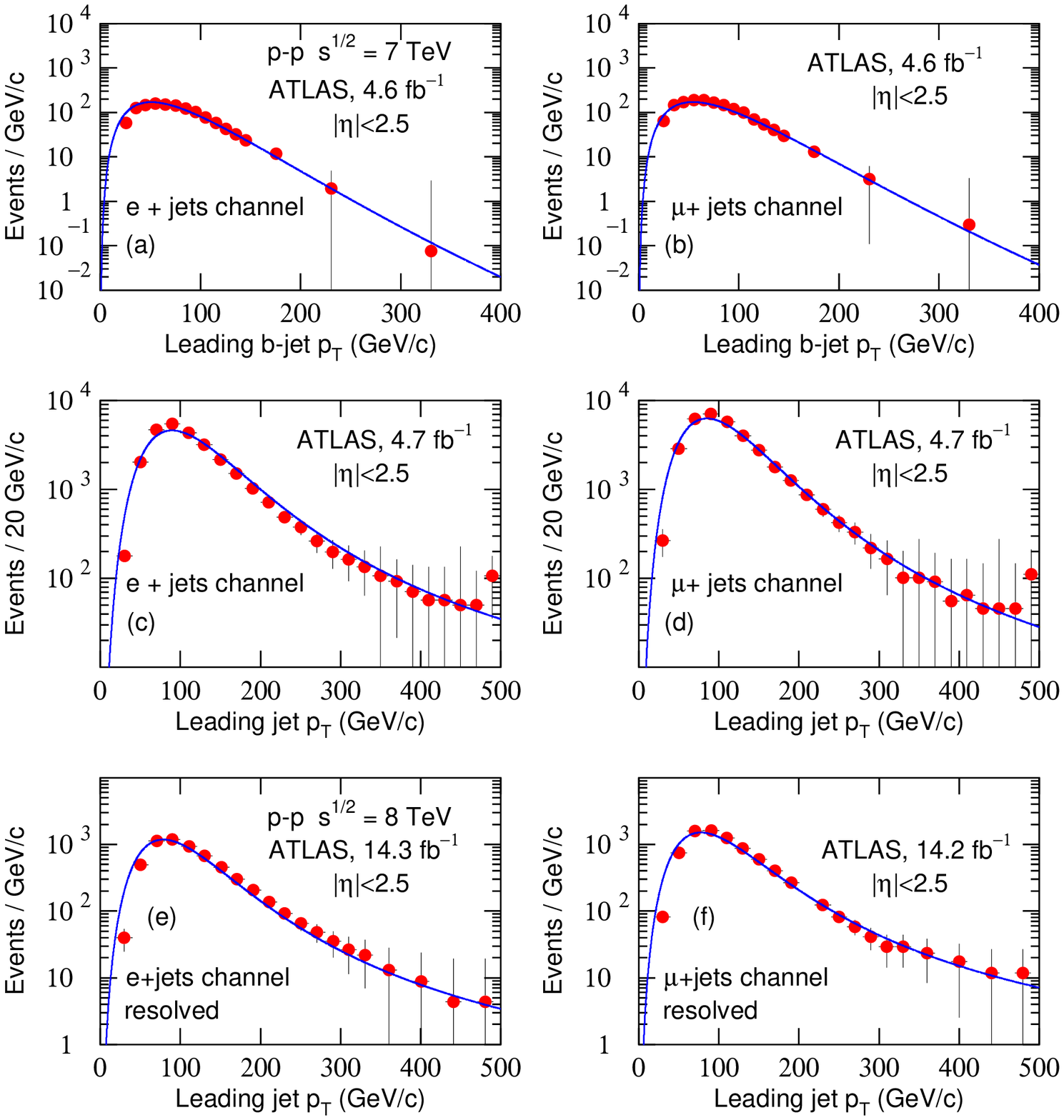}
\end{center}
{\small Figure 4. Transverse momentum spectra of (a)(b) leading
$b$-jets and (c)--(f) leading jets corresponding to the (a)(c)(e)
$e$+jets channel and (b)(d)(f) $\mu$+jets channel in p-p
collisions at (a)--(d) $\sqrt{s}=7$ and (e)(f) 8 TeV. The symbols
are cited from the experimental data measured by the ATLAS
Collaborations~\cite{24,25,26} and the curves are our fitted
results with Eq. (4).}
\end{figure*}

\begin{figure*}[!htb]
\begin{center}
\includegraphics[width=9.0cm]{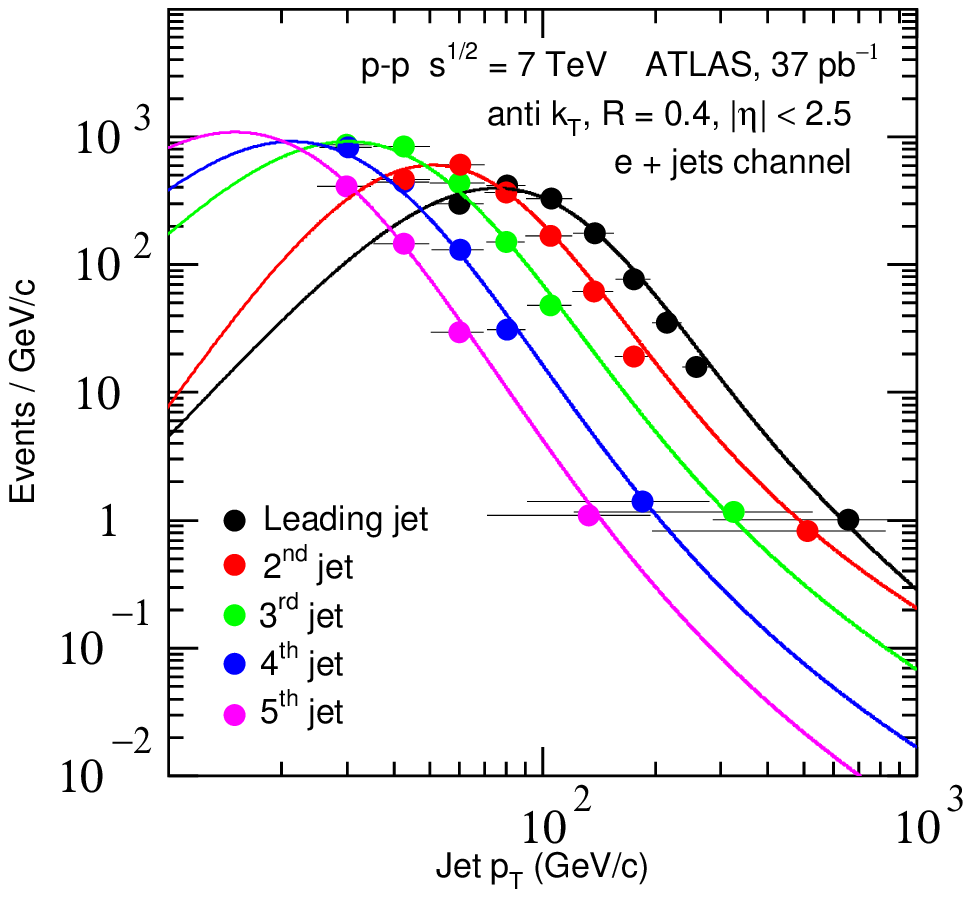}
\end{center}
{\small Figure 5. Transverse momentum spectra of reconstructed
jets with different orders in p-p collisions at $\sqrt{s}=7$ TeV.
The symbols are cited from the experimental data measured by the
ATLAS Collaboration~\cite{27} and the curves are our fitted
results with Eq. (4).}
\end{figure*}

\begin{figure*}[!htb]
\begin{center}
\includegraphics[width=15.0cm]{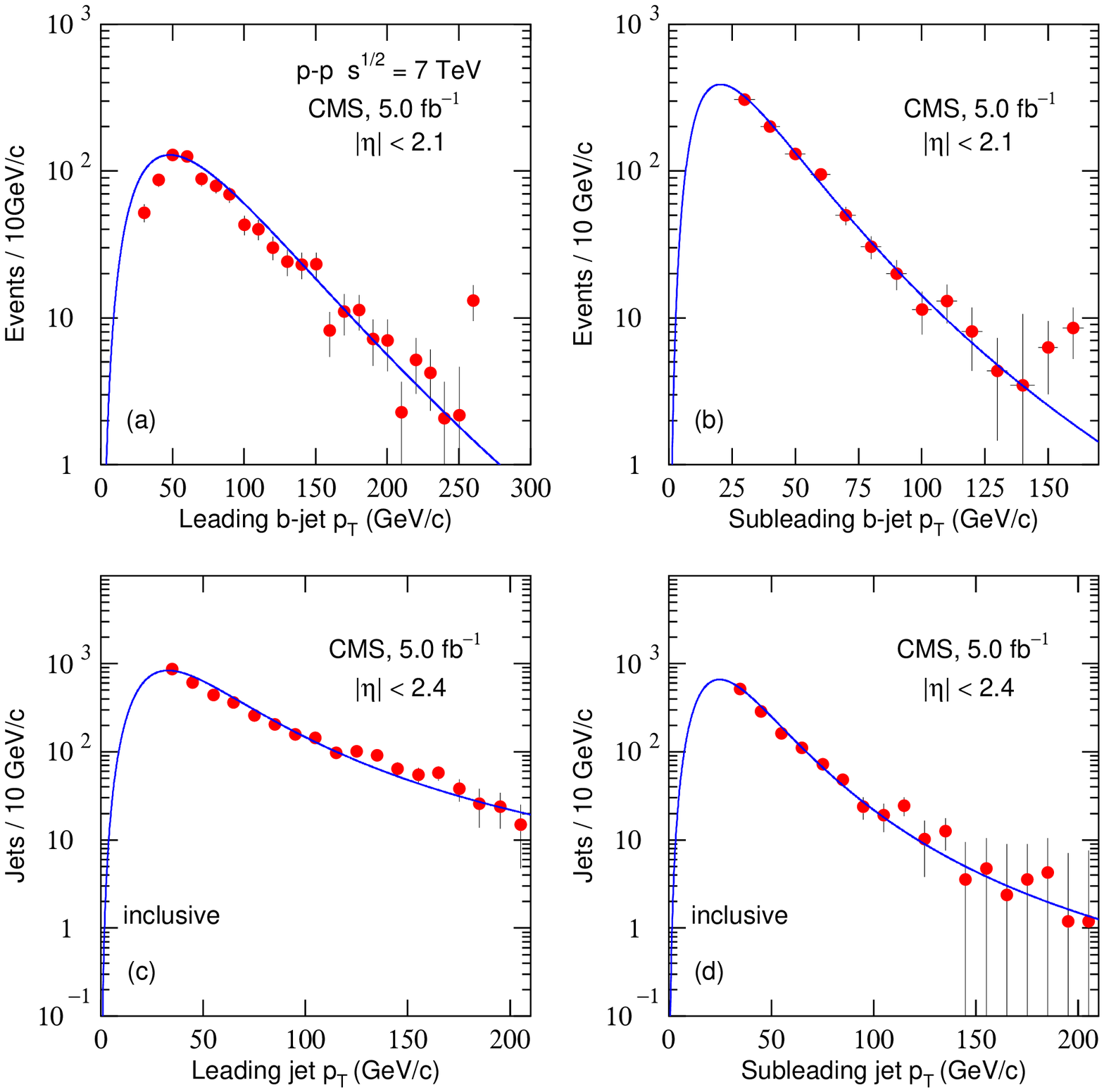}
\end{center}
{\small Figure 6. Transverse momentum spectra of (a) leading
$b$-jets, (b) subleading $b$-jets, (c) leading jets, and (d)
subleading jets produced in p-p collisions at $\sqrt{s}=7$ TeV.
The symbols are cited from the experimental data measured by the
CMS Collaboration~\cite{22,28} and the curves are our fitted
results with Eq. (4).}
\end{figure*}

\begin{figure*}[!htb]
\begin{center}
\includegraphics[width=15.0cm]{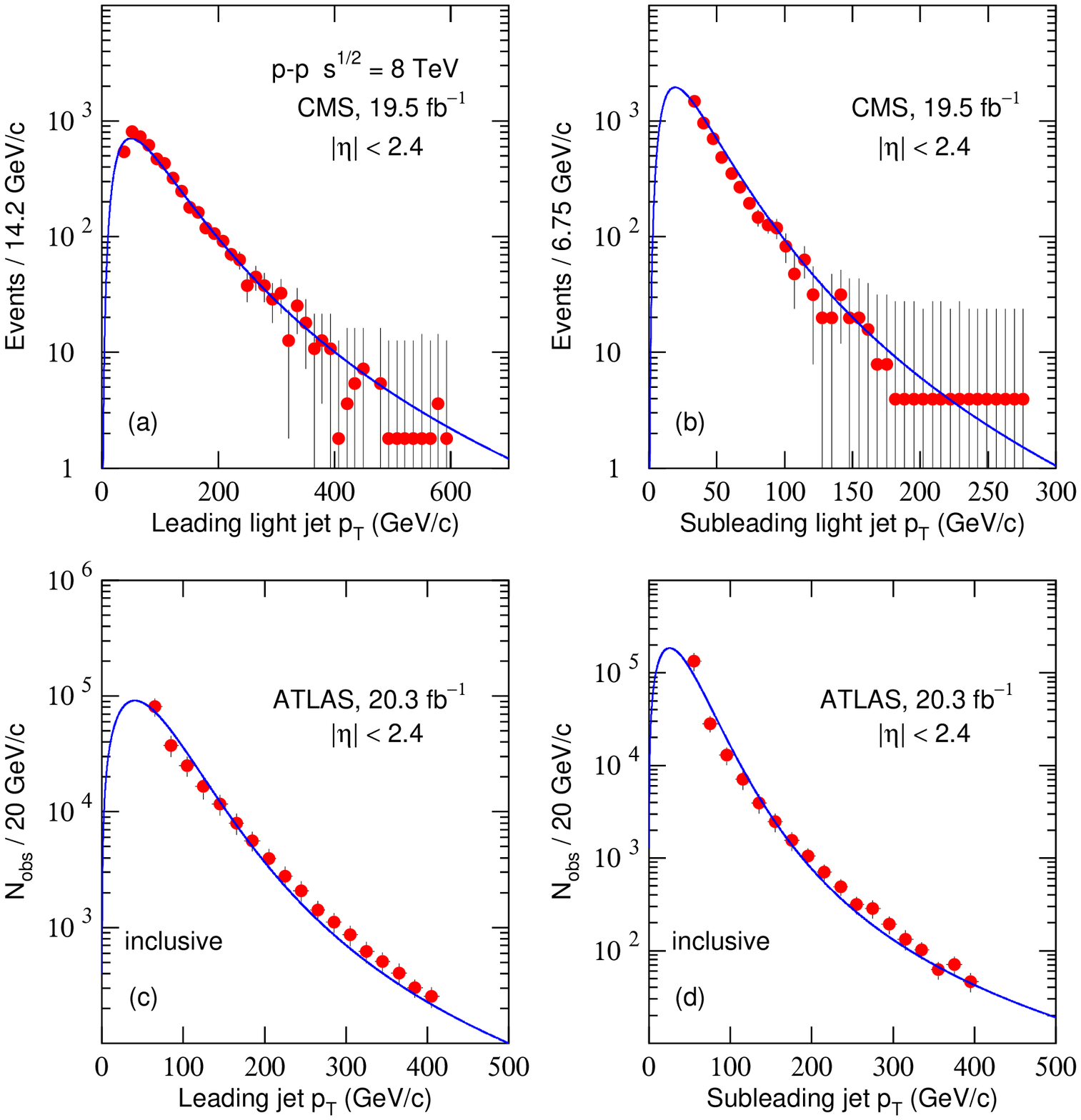}
\end{center}
{\small Figure 7. Transverse momentum spectra of (a) leading light
jets, (b) subleading light jets, (c) leading jets, and (d)
subleading jets produced in p-p collisions at $\sqrt{s}=8$ TeV.
The symbols are cited from the experimental data measured by the
CMS~\cite{29} and ATLAS Collaborations~\cite{30} and the curves
are our fitted results with Eq. (4).}
\end{figure*}

\begin{figure*}[!htb]
\begin{center}
\includegraphics[width=15.0cm]{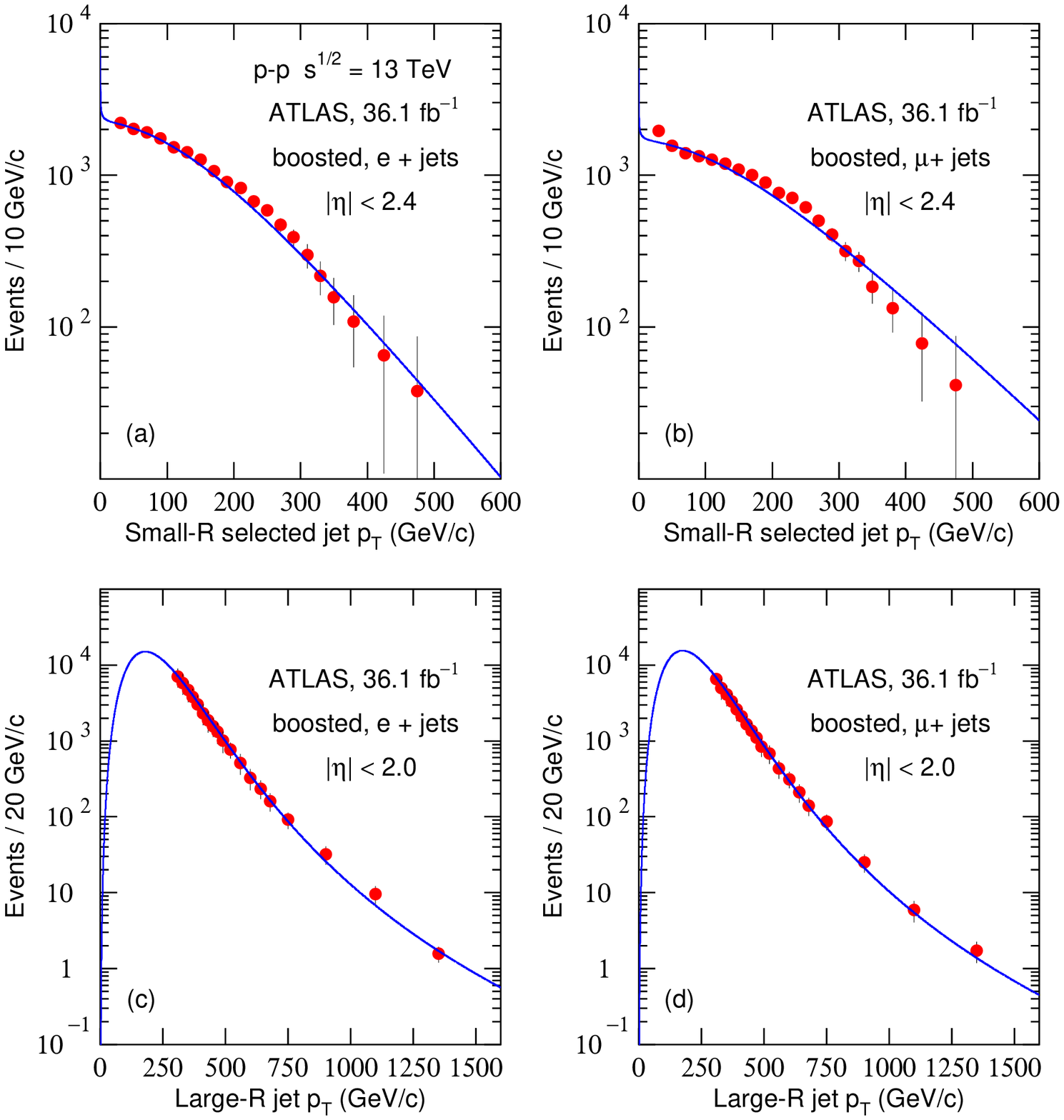}
\end{center}
{\small Figure 8. Transverse momentum spectra of (a)(b) small-R
selected and (c)(d) large-R jets corresponding to the (a)(c)
$e$+jets and (b)(d) $\mu$+jets channels produced in p-p collisions
at $\sqrt{s}=13$ TeV. The symbols are cited from the experimental
data measured by the ATLAS Collaboration~\cite{31} and the curves
are our fitted results with Eq. (4).}
\end{figure*}

\begin{figure*}[!htb]
\begin{center}
\includegraphics[width=15.0cm]{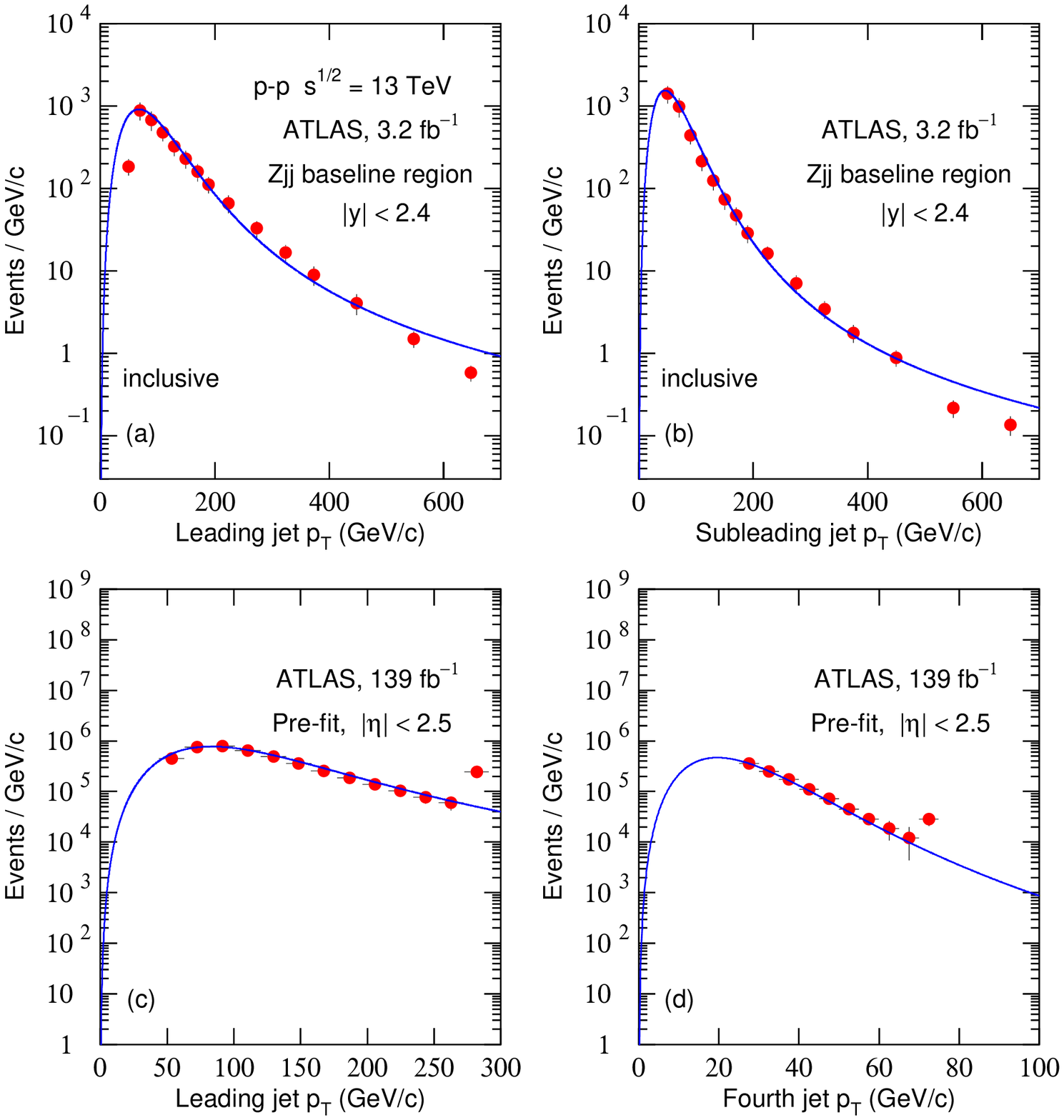}
\end{center}
{\small Figure 9. Transverse momentum spectra of (a) the leading
jets and (b) the subleading jets in $Zjj$ baseline region as well
as (c) the leading jets and (d) the forth jets with pre-fit
produced in p-p collisions at $\sqrt{s}=13$ TeV. The symbols are
cited from the experimental data measured by the ATLAS
Collaboration~\cite{32,33} and the curves are our fitted results
with Eq. (4).}
\end{figure*}

\begin{figure*}[!htb]
\begin{center}
\includegraphics[width=9.0cm]{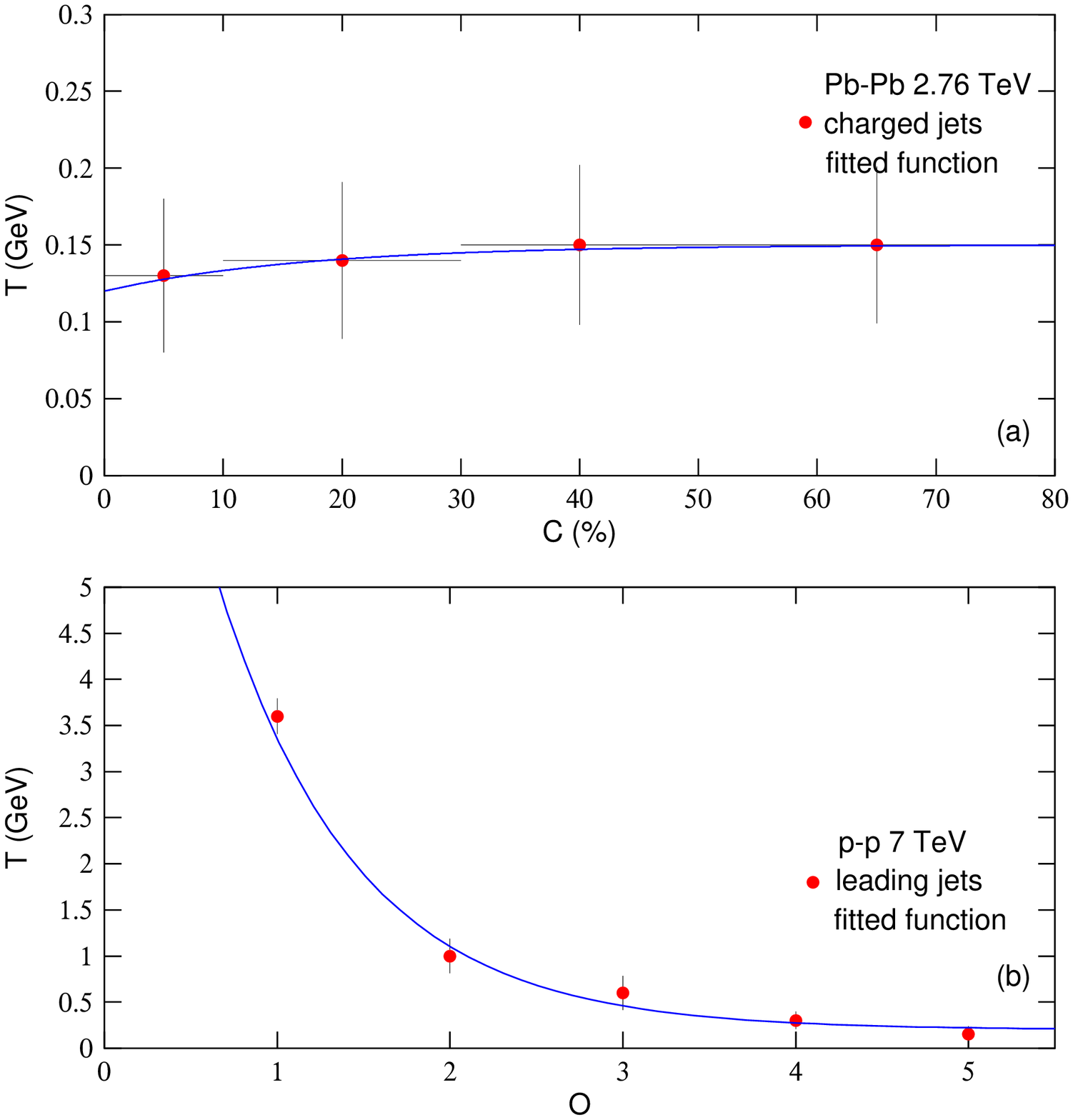}
\end{center}
{\small Figure 10. The relations of (a) the effective temperature
parameter $T$ and the centrality percentage $C$ in Pb-Pb
collisions at $\sqrt{s}=2.76$ TeV, as well as (b) $T$ and the jet
order $O$ in p-p collisions at $\sqrt{s}=7$ TeV. The symbols
represent the values of $T$ obtained from Figures 2 and 5 and
listed in Table 1. The curves are our fitted results with Eqs. (6)
and (7) respectively.}
\end{figure*}

\begin{figure*}[!htb]
\begin{center}
\includegraphics[width=15.0cm]{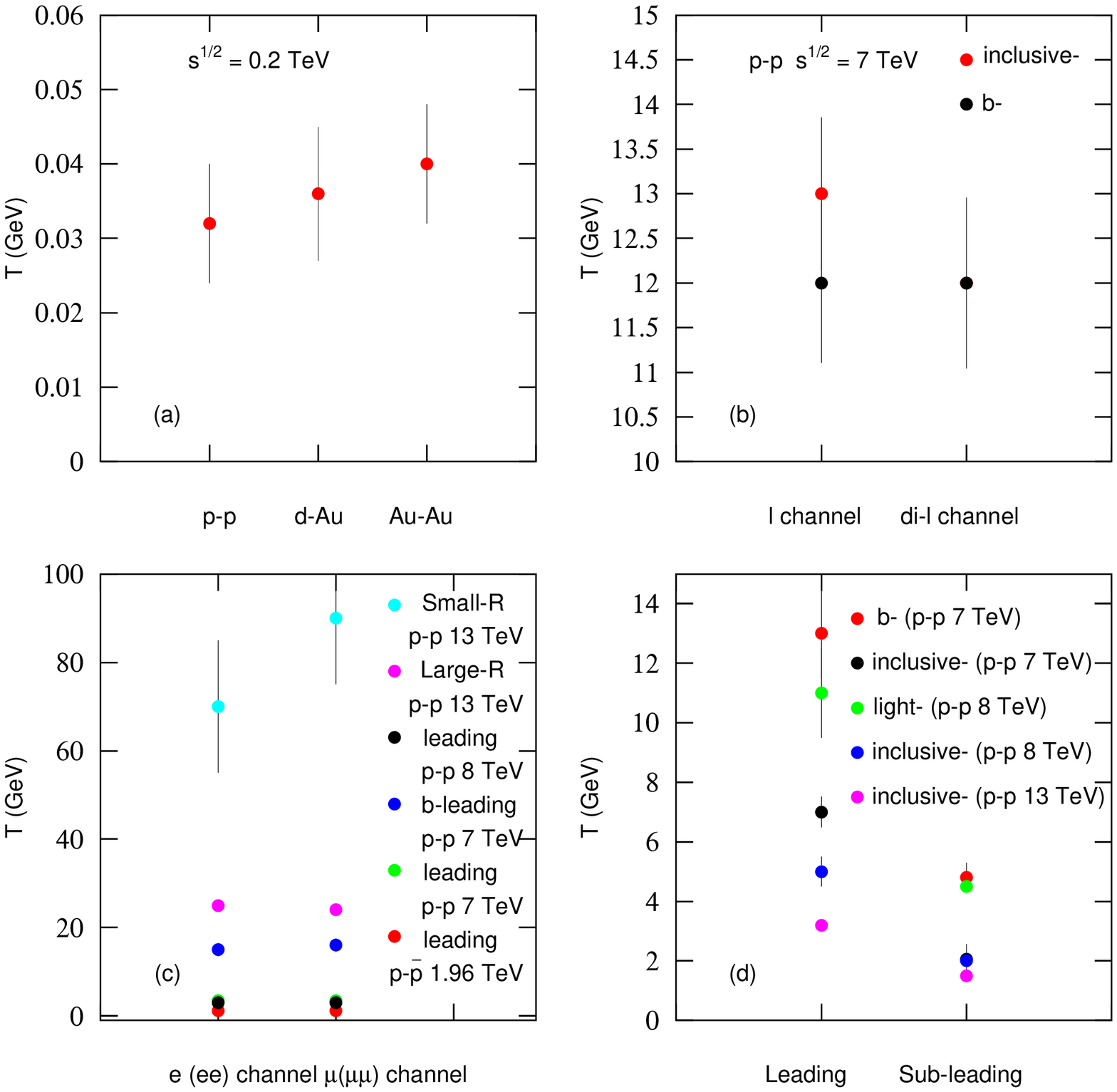}
\end{center}
{\small Figure 11. The relations of the effective temperature
parameter $T$ and (a) the size of interacting system, (b) the
$\ell$ and di-$\ell$ channels, (c) the $e$ ($ee$) and $\mu$
($\mu\mu$) channels, as well as (d) the leading and sub-leading
jets. The symbols represent the values of $T$ obtained from
various figures and listed in Table 1.}
\end{figure*}

{\section{Results and discussion}}

{\subsection{Comparison with data}}

Figure 1 shows the transverse momentum $p_T$ spectra of different
jets produced in (a) p-p, (b) d-Au, and (c) Au-Au collisions with
mid-pseudorapidity (mid-$\eta$, $|\eta|<0.5$ for Figures 1(a) and
1(c) and $|\eta|<0.55$ for Figure 1(b)), as well as in (d) p-p
collisions with non-mid-$\eta$ ($0.8<\eta<1.8$) at
$\sqrt{s_{NN}}=0.2$ TeV, where $N$ and $N_{\rm evt}$ denote the
numbers of jets and events respectively. The symbols are cited
from the experimental data measured by the STAR
Collaboration~\cite{14,15,16}. In Figures 1(a) and 1(c), the high
tower (HT) trigger jets were selected. In Figure 1(a)--(c), the
jet events were selected using a cone radius ($R=0.4$) and
anti-$k_T$ algorithm~\cite{17}, where $k_T$ denotes the transverse
momentum. In Figures 1(b) and 1(c), the data of d-Au and Au-Au
collisions were measured in 0--20\% centrality class. In the
figure, the curves are our fitted results with Eq. (4). In the fit
process, two participant top quarks with constituent mass of 174
GeV/$c^2$ for each one are considered. The values of free
parameters ($n$, $T$, and $a_0$), normalization constant ($N_0$),
$\chi^2$, and number of degree of freedom (ndof) are listed in
Table 1 in which the parameter trend will be analyzed and
discussed in subsection 3.2. One can see that the $p_T$ spectra of
different jets are shown to obey approximately the convolution of
two TP-like functions. The values of mean excitation and expansion
degree (defined by the effective temperature parameter $T$) seem
to not related to the size of collision system in the error range.

Figure 2(a) presents the $p_T$ spectra of fast jets produced in
p-Pb collisions with different centralities at
$\sqrt{s_{NN}}=5.02$ TeV, where $\sigma$ on the vertical axis
denotes the cross section. The $p_T$ spectra of charged jets
produced in Pb-Pb collisions with different centralities at
$\sqrt{s_{NN}}=2.76$ TeV are presented in Figure 2(b), where
$N_{\rm coll}$ on the vertical axis denotes the number of binary
nucleon-nucleon collisions. The symbols are cited from the
experimental data measured by the ALICE
Collaboration~\cite{18,19}. The jet events were selected with a
cone radius ($R=0.2$) and mid-$\eta$ ($|\eta|<0.5$). The curves
are our fitted results with Eq. (4), in which two participant top
quarks are considered. The values of $n$, $T$, $a_0$, $N_0$,
$\chi^2$, and ndof are listed in Table 1. One can see that the
convolution of two TP-like functions describes approximately the
experimental data of the mentioned jets. The effective temperature
parameter $T$ are the same with changing the centrality percentage
in p-Pb collisions. And in Pb-Pb collisions, $T$ increases
slightly with the increase of centrality percentage, i.e. $T$
decreases slightly with the increase of centrality itself.

Figure 3(a) (3(b)) presents the $p_T$ spectrum of leading jets
corresponding to the $Z\rightarrow\mu\mu$ ($Z\rightarrow ee$)
channel in p-$\rm\bar p$ collisions at $\sqrt{s}=1.96$ TeV. Figure
3(c) (3(d)) presents the $p_T$ spectra of jets ($b$-jets)
corresponding to the lepton and dilepton channels in p-p
collisions at $\sqrt{s}=7$ TeV. The symbols are cited from the
experimental data measured by the D0~\cite{20,21}, CMS~\cite{22},
and ATLAS Collaborations~\cite{23}. In Figures 3(a) and 3(b), the
jet events were selected with a cone radius ($R=0.5$) and wide
$\eta$ range ($|\eta|<2.5$). In Figure 3(c) and 3(d), the jet
events were selected with $|\eta|<2.4$ and $|\eta|<2.5$
respectively. The curves are our fitted results with Eq. (4), in
which two participant top quarks are considered for Figures
3(a)--3(c), and two participant bottom quarks with constituent
mass of 4.19 GeV/$c^2$ for each one are considered for Figure
3(d). The values of $n$, $T$, $a_0$, $N_0$, $\chi^2$, and ndof are
listed in Table 1. One can see that the convolution of two TP-like
functions provides an approximate description on the data. The
effective temperature parameter $T$ obtained from the spectra with
the lepton and dilepton channels are almost the same within the
error range.

The $p_T$ spectra of (a)(b) leading $b$-jets and (c)--(f) leading
jets corresponding to the (a)(c)(e) $e$+jets channel and (b)(d)(f)
$\mu$+jets channel in p-p collisions at (a)--(d) $\sqrt{s}=7$ and
(e)(f) 8 TeV are presented in Figure 4. The symbols are cited from
the experimental data measured by the ATLAS
Collaboration~\cite{24,25,26}. The jet events were selected with
$|\eta|<2.5$. The curves are our fitted results with Eq. (4), in
which two participant bottom quarks are considered for Figures
4(a) and 4(b), and two participant top quarks are considered for
Figures 4(c)--4(f). The values of $n$, $T$, $a_0$, $N_0$,
$\chi^2$, and ndof are listed in Table 1. One can see that the
fold of two TP-like functions provides an approximate description
on the data. The effective temperature parameter $T$ from the
$e$+jets channel and the $\mu$+jets channel are almost the same
within the error range.

The reconstructed jet $p_T$ spectra for the leading, $2^{nd}$,
$3^{rd}$, $4^{th}$, and $5^{th}$ order jets in the $e$+jets
channels produced in p-p collisions at $\sqrt{s}=7$ TeV are shown
in Figure 5. The symbols are cited from the experimental data
measured by the ATLAS Collaboration~\cite{27}. The jet events were
selected with a cone radius ($R=0.4$) and wide $\eta$ range
($|\eta|<2.5$). The curves are our fitted results with Eq. (4), in
which two participant top quarks are considered. The experimental
data are approximately fitted with the convolution of two TP-like
functions and the values of related parameters are given in Table
1. One can see that the effective temperature parameter $T$
decreases with the growth of jet order $O$.

Figure 6 presents the $p_T$ spectra of (a) leading $b$-jets, (b)
subleading $b$-jets, (c) leading jets, and (d) subleading jets
produced in p-p collisions at $\sqrt{s}=7$ TeV. The symbols are
cited from the experimental data measured by the CMS
Collaboration~\cite{22,28}. The jet events were selected with
(a)(b) $|\eta|<2.1$ and (c)(d) $|\eta|<2.4$. The curves are our
fitted results with Eq. (4), in which two participant bottom
quarks are considered for Figures 6(a) and 6(b), and two
participant top quarks are considered for Figure 6(c) and 6(d).
The experimental data are approximately fitted with the
convolution of two TP-like functions and the values of related
parameters are given in Table 1. The values of $T$ from the
spectra of leading jets are much larger than those from the
spectra of subleading jets.

Figure 7 displays the $p_T$ spectra of (a) leading light jets, (b)
subleading light jets, (c) leading jets, and (d) subleading jets
produced in p-p collisions at $\sqrt{s}=8$ TeV, where $N_{\rm
obs}$ denotes the number of observation. The symbols are cited
from the experimental data measured by the CMS~\cite{29} and ATLAS
Collaborations~\cite{30}. The jet events were selected with
$|\eta|<2.4$. The curves are our fitted results with Eq. (4), in
which two participant charm quarks with constituent mass of 1.27
GeV/$c^2$ for each one are considered for Figures 7(a) and 7(b),
and two participant top quarks are considered for Figure 7(c) and
7(d). The experimental data are approximately fitted with the
convolution of two TP-like functions and the values of related
parameters are given in Table 1. Once more, the values of $T$ from
the spectra of leading jets are much larger than those from the
spectra of subleading jets.

The $p_T$ spectra of (a)(b) small-R selected and (c)(d) large-R
jets corresponding to the (a)(c) $e$+jets and (b)(d) $\mu$+jets
channels produced in p-p collisions at $\sqrt{s}=13$ TeV are shown
in Figure 8. The symbols are cited from the experimental data
measured by the ATLAS Collaboration~\cite{31}. The curves are our
fitted results with Eq. (4), in which two participant top quarks
are considered. The experimental data are approximately fitted
with the convolution of two TP-like functions and the values of
related parameters are given in Table 1. One can see that the
values of $T$ from the spectra of $e$+jets and $\mu$+jets channels
are almost the same within the error range.

The $p_T$ spectra of (a) the leading jets and (b) the subleading
jets in $Zjj$ baseline region as well as (c) the leading jets and
(d) the forth jets with pre-fit produced in p-p collisions at
$\sqrt{s}=13$ TeV are presented in Figure 9. The symbols are cited
from the experimental data mesaured by the ATLAS
Collaboration~\cite{32,33}. The jet events were selected with
$|y|<2.4$ for Figures 9(a) and 9(b), and $|\eta|<2.5$ for Figures
9(c) and 9(d). The curves are our fitted results with Eq. (4), in
which two participant top quarks are considered. The experimental
data are approximately fitted with the convolution of two TP-like
functions and the values of related parameters are given in Table
1, in which the normalization is very large due to the fact that
it denotes the event number accumulated, but not the cross section
or jet number per event. One can see that the values of $T$ from
the spectra of the leading jets are much larger than those from
the spectra of the subleading and forth jets.

From the above comparison with data, one can see that the data are
approximately fitted with the convolution of two TP-like
functions. In most cases, the quality of the fits is acceptable
due to appropriate $\chi^2$/ndof. In a few cases, the result of
the fits is worthy of further improvement due to large
$\chi^2$/ndof. To improve the result, we need more suitable
function for the component and/or three or more functions in the
convolution. In addition, there are cases in which some parameters
do not change but others suffer significant changes, in that sense
is not possible to conclude about the characteristics of high
energy collisions. Indeed, as an application of the uniform
method, more comparisons with data are needed in future.
Meanwhile, we hope to improve the result of the fits in future.
\\

{\subsection{Parameter trend and discussion}}

Although the comparison with data is not perfect, one can also see
some trends in most cases. To show the trends of main parameters,
Figure 10(a) presents the relation of the effective temperature
$T$ and the centrality percentage $C$ in Pb-Pb collisions at
$\sqrt{s_{NN}}=2.76$ TeV. The symbols represent the values of $T$
obtained from Figure 2 and listed in Table 1. The curve is our fit
by an exponential function
\begin{align}
T=(& -0.03\pm0.01)\exp\bigg(\frac{-C}{17.00\pm2.00}\bigg) \nonumber\\
& +(0.15\pm0.01),
\end{align}
in which $T$ and $C$ are in GeV and \% respectively. One can see
that $T$ increases slightly with the increase of $C$, or $T$ are
almost the same within the error range when $C$ varies. The
relation between $T$ and $C$ renders that QGP formed in central
Pb-Pb collisions has less influence on the jet transport. Or, in
the transport process of jets in QGP in central Pb-Pb collisions,
jets lost less energy.

Figure 10(b) presents the relation of the effective temperature
$T$ and the jet order $O$ in p-p collisions at $\sqrt{s}=7$ TeV.
The symbols represent the values of $T$ obtained from Figure 5 and
listed in Table 1. The curve is our fit by an exponential function
\begin{align}
T=(& 11.00\pm0.10)\exp\bigg(\frac{-O}{0.80\pm0.01}\bigg) \nonumber\\
& +(0.20\pm0.01),
\end{align}
in which $T$ is in GeV. One can see that $T$ decreases with the
growth of $O$. This trend is natural due to the fact that the jet
with high order corresponds to the source with less excitation
degree.

Figure 11 shows the relations of the effective temperature $T$ and
(a) the size of interacting system, (b) $\ell$ and di-$\ell$
channels, (c) $\mu$($\mu\mu$) and $e(ee)$ channels, and (d)
leading and sub-leading jets. The symbols represent the values of
$T$ obtained from the above figures and listed in Table 1. One can
see that $T$ seems to not related to the system size in the error
range. This is in agreement with the conclusion from Figure 10(a)
in which central collisions correspond to large system and
peripheral collisions correspond to small system. In the error
range, different lepton channels show nearly the same effective
temperature, which renders nearly the same excitation degree of
source. At the same time, the values of $T$ from the spectra of
leading jets are much larger than those from the spectra of
subleading jets, which is the same as the conclusion from Figure
10(b).

As a parameter determining the curvature in middle-$p_T$ region
and the extended range in high-$p_T$ region, $n$ is related to the
entropy index $q$ because $n=1/(q-1)$. In most cases, $q\ge1.2$
which is not close to 1 because $n\le5$ which is not large. This
implies that the source of jets does not stay at the equilibrium
state. In a few cases, $n$ is large and $q$ is close to 1. This
happens coincidentally, but not implies that the source of jets
stay at the equilibrium state. This situation is different from
the source of identified particles. Generally, the source of
identified particles stays approximately at the equilibrium or
local equilibrium state. These non-equilibrium and (local)
equilibrium states do not mean a contradiction between the
productions of various jets and identified particles. In fact, the
two types of products are produced at different stages of system
evolution. Generally, various jets are produced at the initial
stage, and the identified particles are produced at the kinetic
freeze-out stage. From the initial to kinetic freeze-out stages,
the collision system evolves quickly from non-equilibrium to
(local) equilibrium states.

An a parameter determining the slope of the curve in low-$p_T$
region, $a_0$ is elastic from negative to positive values. A
negative $a_0$ results in a cocked up distribution and a positive
$a_0$ results in a falling distribution. In many cases, $a_0\neq1$
which means that it is necessary introducing $a_0$ in the
Tsallis--Pareto-type function. Due to the introduction of $a_0$,
the revised Tsallis--Pareto-type function, i.e. the TP-like
function becomes more flexible. The convolution of two or more
TP-like functions is expected to fit more $p_T$ spectra in high
energy collisions. The effective temperature $T$ from the spectra
of leading jets is much larger than that from the spectra of
subleading jets because the leading jets undergone more violent
scattering. However, $a_0$ parameter does not change usually with
the order of jet because $a_0$ determines only the spectra at
low-$p_T$. To determine $T$, the spectra at medium- and high-$p_T$
play main role. Generally, the parameters related to various jets
are not associated with the flow because various jets are produced
at the initial stage where the flow is not formed. The flow is
associated with identified particles which are finally produced at
the kinetic freeze-out stage where the flow is formed.

It is accepted that, regarding central A-A collisions at high
energies, jets lose considerably energy as they propagate through
the hot and dense medium~\cite{34}. Due to the loss of energy, the
$p_T$ spectra are reduced and we obtain seemingly a relative small
$T$ and/or large $n$ (small $q$) comparing with peripheral A-A
collisions. However, due to large error, $T$ in A-A collisions
also shows a nearly invariant trend. In central p-A collisions,
jets do not lose considerably energy and the $p_T$ spectra are not
reduced due to similar small participant volume comparing with
peripheral p-A collisions. This renders that the main parameters
in p-A collisions are independent of centrality, as what appear in
A-A collisions. Mostly, the spectra of jets cited from p-A and A-A
collisions are not under the same condition, it is hard to compare
the parameters directly.

In A-A collisions, the medium effect on the probe of hard process
is commonly referred to as jet quenching. The effect has been
observed via deviations from a well-calibrated baseline
established in the absence of a medium (e.g. in minimum-bias p-p
collisions). The generic strategy is illustrated with the nuclear
modification factor ($R_{AA}$), which evaluates the deviation of a
single particle inclusive spectra away from the baseline. $R_{AA}$
is expected to be one in the absence of medium effect. Jet
measurements in central A-A collisions at high energies have shown
that $R_{AA}$ is $p_T$ dependent and it is smaller than one. For
peripheral A-A collisions $R_{AA}$ approaches to unity and it is
nearly flat (see e.g.~\cite{35}). An analogous analysis in p-A
collisions gives a $R_{pA}$ which is consistent to one (see
e.g.~\cite{18}).

It seems that these features of data on $R_{AA}$ are not captured
by the parameters presented in Table 1. For instance, the
extracted parameters which characterize the spectral shape ($a_0$,
$T$, and $n$) are the same for central A-A collisions and 0-100\%
p-A collisions. Probably this is not a surprise because based on
the reduced $\chi^2$, the proposed function seems not describe all
the data which are analyzed in the paper. This seems to illustrate
that the paper also lacks consistent results. In fact, these
opinions cannot be obtained because the parameters from A-A and
p-A collisions cannot be compared directly. As we know, in Figure
2, the spectra in A-A and p-A collisions are from different kinds
of jets. The same parameters for the two collisions are
coincidental due to the similar trends of data quoted.

Before summary and conclusions, we would like to emphasize the
functions of parameters. As what we discussed in our recent
work~\cite{13}, the power index $a_0$ describes flexibly the
shapes of spectra at low-$p_T$. From negative to positive $a_0$,
the spectra bend from up to down over a $p_T$ range from 0 to 1
GeV/$c$. Correspondingly, the spectra at medium- and high-$p_T$
change higher due to the result of normalization. With the
increase of $T$ and by fixing $a_0$ and $n$, the spectra become
wider. Meanwhile, with the increase of $n$ and by fixing $a_0$ and
$T$, the spectra become narrower. However, the changes with the
increases of $T$ and $n$ cannot be offset. Anyhow, the
introduction of $a_0$ makes the TP-like function more flexible.

We would like to point out that the effective temperature is one
of the reported parameters, however, the kinetic freeze-out
temperature ($T_0$) should be more important to disentangle the
$T_0$ and flow. We have been worked on it for identified particles
in our previous work~\cite{36}. In fact, that paper reports an
increase of $T_0$ with the collision energy, as well as the mass
dependence of the $T_0$. The present work and our previous work
are not contradictory. However, we cannot compare them directly
due to different stages of considerations. The identified
particles are studied at the stage of kinetic freeze-out, while
various jets are produced at the stage of initial collisions. In
addition, different fit functions which correspond to different
``thermometers" are used, which also increases the degree of
difficulty for direct comparisons.
\\

{\section{Summary and conclusions}}

We summarize here our main observations and conclusions.

(a) The transverse momentum $p_T$ spectra of various jets selected
in different conditions and produced in different collisions over
an energy range from 0.2 to 13 TeV are fitted by the convolution
of two TP-like functions, where the TP-like function is a revised
Tsallis--Pareto-type function. The experimental data recorded by
various collaborations are approximately fitted by the mentioned
convolution.

(b) From the fit on the $p_T$ spectra of charged jets produced in
Pb-Pb collisions at $\sqrt{s}=2.76$ TeV with different centrality
intervals, we know that the effective temperature $T$ increases
slightly with increasing the centrality percentage, or $T$ is
almost the same in the error range when the centrality changes.
Meanwhile, $T$ from the spectra of jets in p-p, d-Au, and Au-Au
collisions at 0.2 TeV does not show the size dependence. This is
consistent to the nearly independence of $T$ on centrality.

(c) The values of $T$ from the spectra of leading jets are much
larger than those from the spectra of subleading jets due to the
leading jets undergone more violent scattering. As expected, $T$
extracted from the reconstructed jets produced in p-p collisions
at $\sqrt{s}=7$ TeV decreases with the growth of the jet order. In
addition, $T$ from the lepton and dilepton channels are almost the
same, which means that these jets have common property.

(d) The parameter $n$ determines the curvature in middle-$p_T$
region and the extended range in high-$p_T$ region. Meanwhile, $n$
is related to the entropy index $q$ because $n=1/(q-1)$.
Generally, $n$ is not too large. This means that $q$ is not close
to 1 and the source of jets does not stay at the equilibrium
state. This is different from the source of identified particles
which stays approximately at the equilibrium or local equilibrium
state.

(e) The parameter $a_0$ determines the slope of the curve in
low-$p_T$ region. A negative $a_0$ results in a cocked up
distribution and a positive $a_0$ results in a falling
distribution. Due to the introduction of $a_0$ in the
Tsallis--Pareto-type function, the revised function, i.e. the
TP-like function, becomes more flexible. The convolution of two or
more TP-like functions is expected to have more applications.
\\
\\
{\bf Data Availability}

The data used to support the findings of this study are included
within the article and are cited at relevant places within the
text as references.
\\
\\
{\bf Ethical Approval}

The authors declare that they are in compliance with ethical
standards regarding the content of this paper.
\\
\\
{\bf Disclosure}

The funding agencies have no role in the design of the study; in
the collection, analysis, or interpretation of the data; in the
writing of the manuscript; or in the decision to publish the
results.
\\
\\
{\bf Conflict of Interest}

The authors declare that there are no conflicts of interest
regarding the publication of this paper.
\\
\\
{\bf Acknowledgments}

The work of the first author (Y.M.T.) was supported by Shanxi
University. The work of the second author (P.P.Y.) was supported
by the China Scholarship Council (Chinese Government Scholarship)
under Grant No. 202008140170 and the Shanxi Provincial Innovative
Foundation for Graduate Education under Grant No. 2019SY053. The
work of the third author (F.H.L.) was supported by the National
Natural Science Foundation of China under Grant Nos. 11575103 and
11947418, the Scientific and Technological Innovation Programs of
Higher Education Institutions in Shanxi (STIP) under Grant No.
201802017, the Shanxi Provincial Natural Science Foundation under
Grant No. 201901D111043, and the Fund for Shanxi ``1331 Project"
Key Subjects Construction.
\\
\\
\end{multicols}
\end{document}